# On The Vibrational Flux in Bounded Atoms

N. A. M. dos S. Caturello

*Abstract*

In this paper we derived a model based on general assumptions and allowed us to derive some important thermodynamic functions that are time-dependent, also we could see the behavior of these functions by surfaces. The model is based on independent movements that couple and construct a flux, which makes the system as a whole not to be independent at all.

*Introduction*

Structural patterns are of much interest in nowadays science [1, 2]. The number of solid state structures that reach fractal dimensions give us a new and exciting field of chemistry and physics to be worked with. This model deals in a very heuristic way with the H-Theorem. Once it has been asserted the conceptual linkage between perturbed oscillators and this theorem [3] we assumed the simplest time-energy correlation that gave us a nonlinear function behavior. Our approach is based on a single time-dependent averaged harmonic oscillator and from it we derived a partition function also time-dependent. Our derivation is rather based on non-perturbational methods. In this sense it can be applied to systems undergoing flux regimes in which it is necessary to take into account any vibrational flux.

*The Model*

This model was developed for a system that contains atoms in a spatial pattern. For that we should consider into a valid approximation that its individual energies summed are equal to its total energy. As a first step we may consider the harmonic oscillator to describe any variation that is able to cause an energetic flux. Therefore we begin the derivation of the partition function based on an energetic ladder which levels are separated by an energy we name as *hv*.

$$q^{vib} = \sum_n g_n e^{-\beta h v_n} \quad (1)$$

Where $q^{vib_\sigma}$ is the vibrational partition function, $g_n$ is the degeneracy of each energetic level, and the exponential term takes into account the energy of these levels. As a convention, we take the first energetic level as being arbitrarily equal zero, by this, we have:

$$q^{vib} = (1 + g_n e^{-\beta h v_n} + g_n e^{-2\beta h v_n} + \ldots) \quad (2)$$

Considering that the degeneracy is unity then we have:

$$q^{vib} = (1 + e^{-\beta h v_n} + e^{-2\beta h v_n} + \ldots)$$
$$q^{vib} = 1 + e^{-\beta h v_n}(1 + g_n e^{-\beta h v_n} + \ldots)$$
$$q^{vib}(1 - e^{-\beta h v_n}) = 1$$
$$q^{vib} = \frac{1}{(1 - e^{-\beta h v_n})} \quad (3)$$

As we have considered at first we can write the variation of the average vibration energy of flux in function of time as being:

$$\frac{d<E^{vib_\sigma}>}{dt} = -\frac{\partial}{\partial t}\left(\frac{\partial \ln q^{vib_\sigma}}{\partial \beta}\right) = \frac{\partial(\sum_j P_j E_j^{vib_\sigma})}{\partial t} \quad (4)$$

The terms in (4) are expected not to be null if the system is not in equilibrium. In a general sense, we can say that the average energy we have in the system equals the term *nRT*, where *R* is the universal constant of the gases, *n* is the number of moles of the substance and *T* is the temperature. As we can deduce, the partition function that takes into account time as an explicit variable can be written as:

$$Ln \; q^{vib_\sigma}(t,T,N) = \int_{t_0}^{t} \int_{\beta_0}^{\beta} \frac{N}{\beta t}\left(1 - e^{-\beta h \nu}\right) dt \, d\beta \quad (5)$$

We now integrate and give the explicit form of this equation, the one that follows:

$$Ln \; q^{vib_\sigma}(t,T,N) = NLog\left(\frac{t}{t_0}\right)[Ei(-\beta h\nu) - Log(\beta)]_{\beta_0}^{\beta} \quad (6)$$

Where the "Ei" represents the exponential integral function and *N* represents $nN_A$, being $N_A$ the Avogadro's number. But if we say that:

$$Log\left(\frac{t}{t_0}\right) = \Gamma \quad (7)$$

We can write (vi) in this form:

$$Ln \; q^{vib_\sigma}(t,T,N) = N\Gamma[Ei(-\beta h\nu) - Log(\beta)]_{\beta_0}^{\beta} \quad (8)$$

Then we have that:

$$q^{vib_\sigma}(t,T,N) = e^{N\Gamma[Ei(-\beta h\nu) - Log(\beta)]_{\beta_0}^{\beta}} \quad (9)$$

We can use for the above equation an initial temperature condition comparable to the lowest energetic level of the harmonic oscillator, and so it will result in:

$$T_0 = \frac{h\nu}{2k} \quad (10)$$

In this sense, (8) takes the form below:

$$q^{vib_\sigma}(t,T,N) = e^{N\Gamma[Ei(-\beta h\nu) + Log(2/h\nu) - Log(\beta) - \theta]} \quad (11)$$

Where θ is simply the number Ei(2). The surface of (11) is shown below:

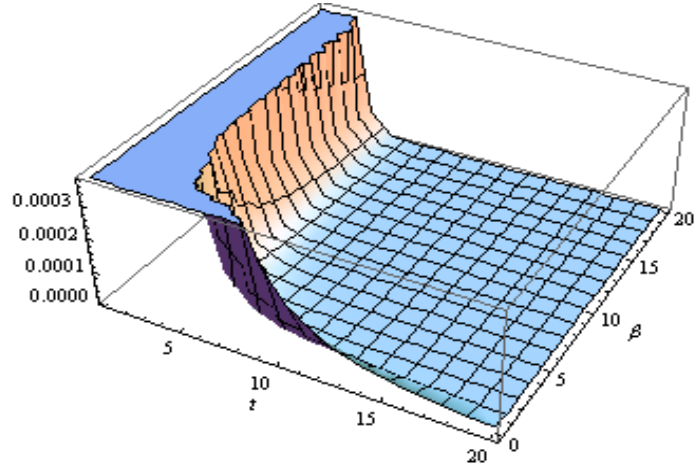

**Figure 1:** The behavior of $q^{vib_\sigma}$ in function of time (t) and β (both in arbitrary units).

In the surface above all the other values of physical features in Eq. (11) were used as unity. Only the values of time (t) and β varied in arbitrary values. With the partition function we were able to derive the equations of other thermodynamic properties of the system. At first we derived the Helmholtz energy of flux ($F^{vib_\sigma}$) the equation is shown below:

$$F^{vib_\sigma}(t,T,N) = -\frac{N\Gamma[Ei(-\beta h\nu)+Log(2/h\nu)-Log(\beta)-\theta]}{\beta} \quad (12)$$

The surface of this equation is shown below:

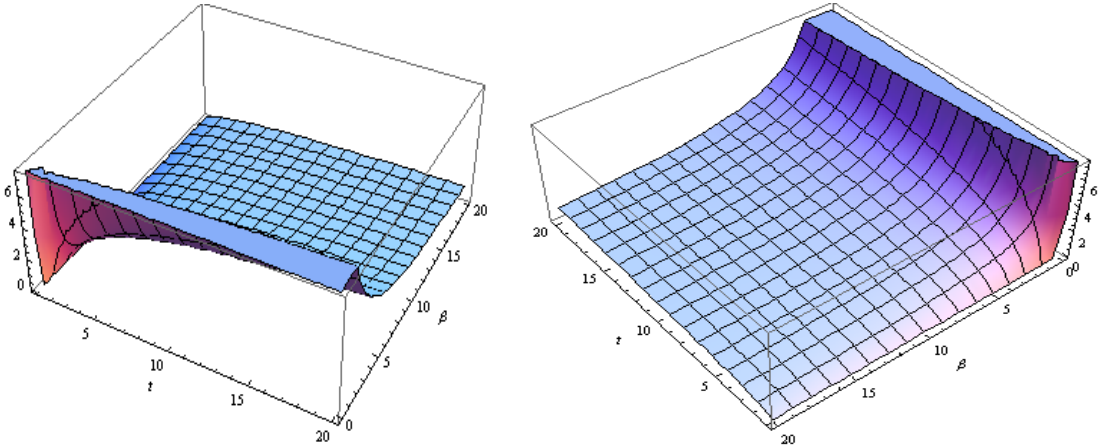

**Figure 2:** Two positions showing the variation in the Helmholtz energy (same considerations about the units made in Fig. 1).

From (12) we were able to derive an expression for entropy of flux, and it is shown below:

$$S^{vib_\sigma}(t,T,N) = N\Gamma k[Ei(-\beta h\nu) - e^{-\beta h\nu} - Log(\beta) + 1 + \theta + Log(2/h\nu)] \quad (13)$$

With the same unitary considerations we made in Fig (1) and Fig (2) we proceeded on plotting the surface of Eq. (13), which yielded the surface displayed below:

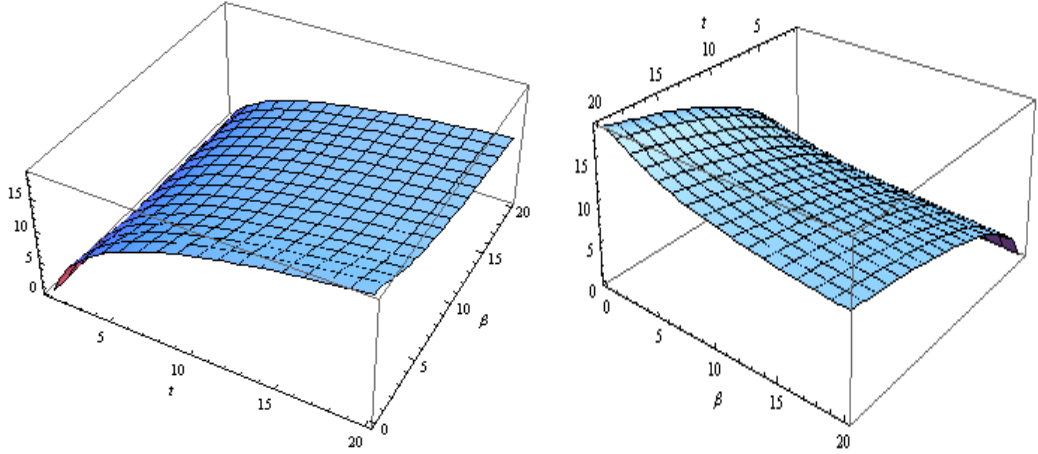

**Figure 3:** The surface of $S^{vib_\sigma}$ in function of time (t) and β.

We could finally derive an equation for $G^{vib_\sigma}$, counting and it is shown below:

$$G^{vib_\sigma}(t,T,N) = -\frac{n\text{r}[Ei(-\beta h\nu)+Log(2/h\nu)-Log(\beta)-\theta]}{\beta} \quad (14)$$

The behavior of the surface of Eq. (14) resembles the one of Fig. 2, under the same unitary considerations. We also derived an equation for the calorific capacity:

$$C_v^{vib_\sigma}(t,T,N) = 3N\text{r}\left[k - \left(\frac{h\nu.e^{-\beta h\nu}}{T} + ke^{-\beta h\nu}\right)\right] \quad (15)$$

The number three is due to Eq. (15) because we consider here a three-dimensional space. The behavior of the surface is shown below:

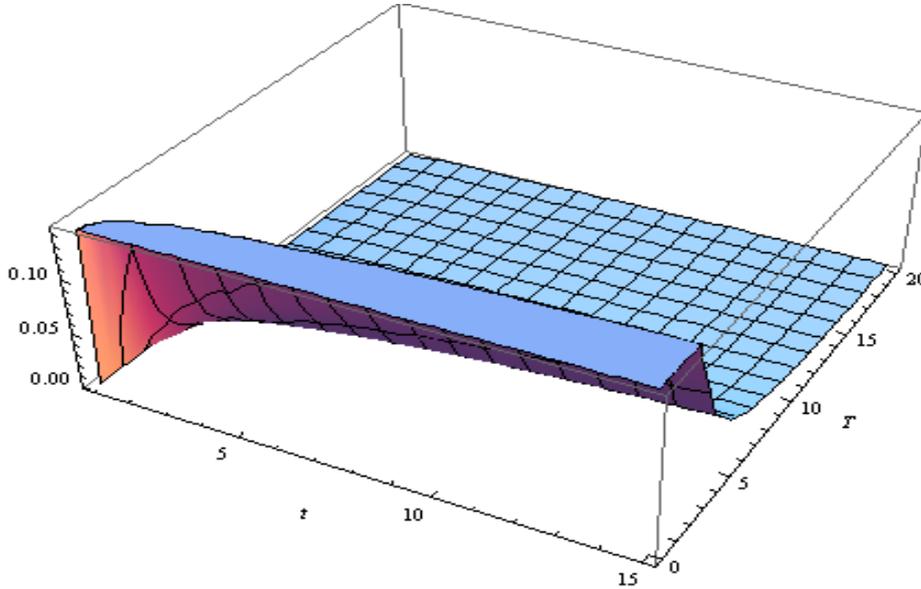

**Figure 4:** Heat Capacity of flux (same unitary considerations made in Fig 1).

*Conclusions*

We see by the behavior of Figs. (1), (2), and (3) an expected variation in function of time and T, what gave to our model physical reliability. The possible explanation of why the heat capacity decreases monotonically with the temperature is that it is easier for a system at high temperatures to receive and/or pass through the energy given to it. Furthermore at T=0 the model predicts a non-null heat capacity, and we interpreted it as the potential energy which arises from intrinsic localized vibrational modes, even though the model used was the one of a harmonic oscillator.

The author has a lot of people to thank, but in special his great friend Lívia and his parents.